\documentclass[aps,prl,reprint,twocolumn,superscriptaddress]{revtex4-1}
\usepackage{amsmath}
\usepackage{amsfonts}
\usepackage{amssymb}
\usepackage{amsthm}
\usepackage{graphicx}
\usepackage{CJK}
\usepackage{times}
\usepackage{mathptmx}
\usepackage[colorlinks, linkcolor=blue, urlcolor=blue, anchorcolor=blue, citecolor=blue]{hyperref}
\usepackage{mathtools}
\usepackage{lipsum}
\usepackage{array}
\usepackage[table]{xcolor}
\definecolor{mygray}{gray}{.9}
\usepackage{empheq}

\begin{document}

\title{Survey of 360$^{\circ}$ domain walls in magnetic heterostructures: topology, chirality and current-driven dynamics}

\author{Mei Li}
\affiliation{Physics Department, Shijiazhuang University, Shijiazhuang, Hebei 050035, People's Republic of China}
\author{Jie Lu}
\email{jlu@hebtu.edu.cn}
\affiliation{College of Physics and Hebei Advanced Thin Films Laboratory, Hebei Normal University, Shijiazhuang 050024, People's Republic of China}

\date{\today}

\begin{abstract}
Chirality and current-driven dynamics of topologically nontrivial 360$^{\circ}$
domain walls (360DWs) in magnetic heterostructures (MHs) are systematically investigated.
For MHs with normal substrates, the static 360DWs are N\'{e}el-type with no chirality.
While for those with heavy-metal substrates, the interfacial Dzyaloshinskii-Moriya interaction (iDMI)
therein makes 360DWs prefer specific chirality.
Under in-plane driving charge currents, as the direct result of ``full-circle" topology 
a certain 360DW does not undergo the ``Walker breakdown"-type
process like a well-studied 180$^{\circ}$ domain wall as the current density increases.
Alternatively, it keeps a fixed propagating mode (either steady-flow or precessional-flow, depending
on the effective damping constant of the MH) until it collapses or changes to other types of solition
when the current density becomes too high.
Similarly, the field-like spin-orbit torque (SOT) has no effects on the dynamics of 360DWs, 
while the anti-damping SOT has.
For both modes, modifications to the mobility of 360DWs by iDMI and anti-damping SOT are provided.

\end{abstract}

\maketitle


\section{I. Introduction}\label{Section_introduction}
The invention and great development of non-volatile magnetic nanodevices 
have led to a profound revolution in the information industry\cite{Leeuw_RPP_1980,Bauer_RMP_2005,Klaui_JPCM_2008}. 
In these nanodevices, various magnetic solitons or magnetic domains they separate
play the roles of 0 and 1 in binary world. 
In wide magnetic nanostrips, 
skyrmions/antiskyrmions\cite{Boni_Science_2009,Nagaosa_Nature_2010,Hoffmann_PhysRep_2017,XiB_NanoLett_2019,ZhangXC_JPCM_2020},
bimerons\cite{Ezawa_PRB_2011,Batista_PRB_2015,Tretiakov_PRB_2019,ShenLC_PRL_2020,ZhangXC_PRB_2020} and so on
are two-dimensional (2D) isolated 
topologically nontrivial magnetic solitions surrounded by connected domains with uniform orientation.
Under the standard definition of 2D topological charge
\begin{equation}\label{2D_winding_number_definition}
\mathcal{W}_{\mathrm{2D}}(\mathbf{m})=\frac{1}{4\pi}\int_{\mathbb{R}^2}\mathbf{m}\cdot\left(\frac{\partial\mathbf{m}}{\partial x}\times\frac{\partial\mathbf{m}}{\partial y}\right)\mathrm{d}(x,y),
\end{equation}
in which $\mathbf{m}$ is a $\mathbb{R}^3$ unit magnetization field locating on the $(x,y)$ plane,
these solitions have an integer $\mathcal{W}_{\mathrm{2D}}$ and they themselves are the information carriers.
While in narrow enough nanostrips which are quasi-one dimensional (Q1D) systems, 
the most studied magnetic solitons are the 1D (N\'{e}el or Bloch) 180$^{\circ}$
domain walls (180DWs) bearing 1/2 1D topological charge, which is defined as
\begin{equation}\label{1D_winding_number_definition}
\mathcal{W}_{\mathrm{1D}}(\mathbf{m})=\frac{1}{2\pi}\int_{\mathbb{R}}\left(m_1\frac{\partial m_2}{\partial\rho}-m_2\frac{\partial m_1}{\partial\rho}\right)\mathrm{d}\rho,
\end{equation}
where $m_{1,2}$ are magnetization components in the wall plane
and the Q1D systems are supposed to extend in $\rho-$direction.
180DWs separate two opposite oriented domains whose orientations can be defined as 0 and 1, 
meantimes the wall motion leads to the transformation of information.
Since the famous Walker analysis\cite{Slonczewski_1972}, tremendous progress has been made on statics and
dynamics of 180DWs driven by various external stimuli\cite{Science_284_468_1999,jlu_EPL_2009,PRL_104_037206_2010,Berger_PRB_1996,Slonczewski_JMMM_1996,PRL_92_207203_2004,YanPeng_PRL_2011,YanPeng_PRL_2012,PRL_113_097201_2014,PRB_90_014414_2014,jlu_PRB_2016,jlu_SciRep_2017,jlu_Nanomaterials_2019,jlu_PRB_2019,jlu_PRB_2020}.
The corresponding results have laid the foundation for many mature commercial and developing 
magnetic nanodevices. 

Interestingly, even in Q1D systems we also have some kinds of isolated magnetic solitions which
have integer $\mathcal{W}_{\mathrm{1D}}$.
Among them, the simplest ones are the so-called 360$^{\circ}$ domain walls (360DWs)
in which the magnetization rotates over one full circle across the intermediate region thus bearing $W_{\mathrm{1D}}=\pm 1$.
In the beginning of 1960s, 360DWs were first found to appear in the magnetization reversal
process of thin films and their existence seem to be a nuisance since they may complicate the reversal process\cite{Smith_JAP_1962,Cohen_JAP_1963,Wade_PhilMag_1964}.
However, studies in the past three decades revealed that 360DWs themselves in 2D magnetic films 
have more interesting physics\cite{Puchalska_JMMM_1991,Liedke_JAP_2006,Muratov_JAP_2008,Dean_JAP_2011,Oshea_JPDAP_2015,Bedanta_PRB_2018}.
Now we know that 360DWs in lower dimensional systems, such as
nanorings\cite{Ross_2011_APL,Aidala_Nanotechnology_2011,Aidala_JAP_2012,Gonzalez_PRB_2013,Aidala_JAP_2014,Muratov_JAP_2015}
and nanostrips\cite{Kubetzka_PRB_2003,Ross_APL_2012,Gonzalez_APL_2013,Ross_NJP_2016,Aidala_AIPAdvances_2016_a,Aidala_AIPAdvances_2016_b},
can be qualified candidates to store and process information in magnetic nanodevices
due to its ``full-circle" topology.
From the viewpoint of application, the energy barrier of nucleating a 360DW in
single-domain nanorings or nanostrips is much lower than a 180DW since in the latter case one should reverse 
the magnetic moments in entire half.
Also in many cases, a 360DW emerges from the combination of two neighboring 180DWs with opposite polarity
due to the long-range magnetostatic interaction or external magnetic fields.

From the beginning of this century, a series of analytical works focus on the question 
whether 360DWs are genuine stable magnetization textures or just long-lived metastable states\cite{Muratov_JAP_2008,Slastikov_Royal_2005,Muratov_J_Comput_Phys_2006}.
For 2D ferromagnetic (FM) films, the main results are as follows: 
(i) the magnetostatics is crucial for the existence of 360DWs; 
(ii) if the long-range component of magnetostatics is neglected, 
an in-plane external field must be applied to stabilize a 1D front of 360DW whose energy 
is independent of wall orientation\cite{Muratov_JAP_2008}. 
As the films fade into narrow enough nanostrips, changes in boundary conditions 
further require that the external field should align with the easy axis to guarantee the existence of 360DWs.
In recent device applications, narrow FM metallic nanostrips often serve as the central components of 
magnetic heterostructures (MHs) with heavy-metal (HM) substrates.
Then the effects of interfacial Dzyaloshinskii-Moriya interaction (iDMI)\cite{Dzyaloshinsky,Moriya}
therein to the chirality preference of 360DWs need to be clarified.

Once nucleated, 360DWs in MHs can be driven by certain external stimuli.
First, external magnetic fields along easy axis can not finish this job.
This can be understood by our roadmap of field-driven domain wall motion
since the Zeeman energy densities in the two domains on both sides of 360DWs
are the same\cite{jlu_EPL_2009}. 
Then, current-induced motion of 360DWs becomes the next choice.
Indeed, it is the most common way to implement and manipulate in real MHs.
Numerical investigations on this issue have been widely preformed in the past decade\cite{Ross_PRB_2010,LiuQingfang_PhysicaB_2012,Gonzalez_PRB_2013,Ross_APL_2013,LiuQingfang_JMMM_2013,HuJingGuo_AIPAdvances_2015,Xibin_srep_2017}.
Alternatively, there are few analytical studies due to
the complexity from the coexistence of iDMI, spin-transfer torque (STT) and spin-orbit torque (SOT) therein.
In this paper, by adopting the Lagrangian-based collective coordinate models (LB-CCMs) and adequate
wall ansatz, the current-driven
dynamics of 360DWs is systematically explored which constitutes the second part of this work.

The paper is organized as follows.
First, the magnetic Lagrangian and dissipation functional of MHs are introduced in Sec. II.
Both perpendicular magnetic anisotropy (PMA) and in-plane magnetic anisotropy (IPMA)
for the central FM metallic layers are considered.
Then in Sec. III.A we define the proper ansatz for 1D topologically nontrivial 360DWs 
and then introduce three typical candidates. 
After integrating over the long axis of MHs, 
a set of unified dynamical equations is obtained in Sec. III.B and serves as the startpoint of our work.
In Sec. III.C, chirality preference of 360DWs selected by iDMI is investigated.
After that, the propagation mode of 360DWs under in-plane currents are systematically explored in Sec. III.D.
Also, for both modes the effects of iDMI and SOT to the dynamics of 360DWs are analytically calculated.
Finally discussions and concluding remarks are provided in Sec. IV and V, respectively.

\section{II. Formulism}\label{Section_Lagrangian}
\begin{figure} [h] 
	\centering
	\includegraphics[width=0.45\textwidth]{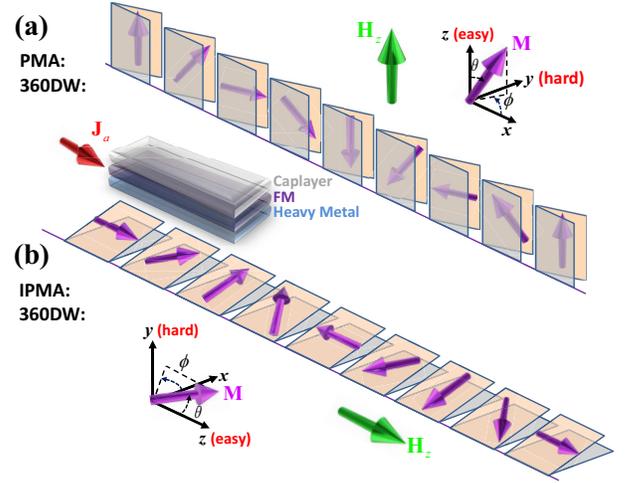}
	\caption{(Color online) Sketch of a MH in which a 360DW
		is formed in its central FM metallic layer with (a) PMA and (b) IPMA.
		A typical MH is composed of a three-layer structure: 
		a HM substrate, a central FM metallic layer and a normal caplayer.
		The corresponding coordinate system is depicted at the up-right and bottom-left corners in the 
		respective subfigure. 
		In each case, the easy (hard) axis lies in the $z(y)-$ direction. 
		An external magnetic field $\mathbf{H}_z=H_z\mathbf{e}_z$ is applied to guarantee the existence of 
		360DWs. When in-plane charge current $\mathbf{J}_a=j_a\mathbf{e}_{\rho}$ is applied,
		magnetization vectors will be driven to tilt from their static locations 
		meantime the 360DW will be driven to propagate along the long axis.
		Gray (orange) planes describe the planar $\varphi-$distribution 
		of static (dynamical) magnetization texture. }\label{fig1}
\end{figure} 
A MH under consideration is shown in Fig. \ref{fig1},
which is composed of three layers: a HM substrate,
a central FM metallic layer and a normal caplayer.
We suppose that the MH is long and narrow enough so that it can be viewed as a Q1D system
extended in the long axis.
For central FM layers with PMA (which will be referred to as ``PMA sytems"), 
the easy axis lies in $z-$axis (out-of-plane normal), the long axis of MH is along
$x-$axis and $\mathbf{e}_y=\mathbf{e}_z\times\mathbf{e}_x$ being the hard axis.
While for central FM layers with IPMA (named as ``IPMA sytems"),
the easy axis coincides with long axis and is defined as $z-$axis,
the hard axis is along out-of-plane normal and denote as $y-$axis,
at last $\mathbf{e}_x=\mathbf{e}_y\times\mathbf{e}_z$.
In these two coordinate systems, the crystalline aisotropy energy density
for PMA and IPMA systems shares the same form.
However, the iDMI should be treated carefully since it is determined by the out-of-plane normal 
component of the magnetization vector.
By setting ``$\mathbf{n}$" as the out-of-plane normal, the iDMI energy density can be written as\cite{Bogdanov_JMMM_1994}
\begin{equation}\label{E_iDMI_expression_general}
\mathcal{E}_{\mathrm{iDMI}}=D_{\mathrm{i}}\left\{\left[\mathbf{m}\left(\mathbf{r}\right)\cdot\mathbf{n}\right]\nabla\cdot\mathbf{m}\left(\mathbf{r}\right)-\left[\mathbf{m}\left(\mathbf{r}\right)\cdot\nabla\right]\left[\mathbf{m}\left(\mathbf{r}\right)\cdot\mathbf{n}\right]\right\},
\end{equation}
where $D_{\mathrm{i}}$ is the iDMI strength and
$\mathbf{m}\left(\mathbf{r}\right)$ is the unit magnetization vector at position $\mathbf{r}$.
Accordingly, the total magnetic energy density $\mathcal{E}_0$ includes
the exchange, crystalline anisotropy, magnetostatic, Zeeman and iDMI energies. 
In narrow enough strips, most of the magnetostatic energy can be described by local
quadratic terms of $M_{x,y,z}$ by means of three average demagnetization factors $D_{x,y,z}$\cite{jlu_PRB_2016}.
In addition, for Q1D systems $\nabla\equiv \frac{\partial}{\partial \rho} \mathbf{e}_{\rho}$ in which
$\rho=x (z)$ for PMA (IPMA) systems. Thus one has
\begin{equation}\label{E_total}
\begin{split}
\mathcal{E}_0[\mathbf{m}]&= A\left(\frac{\partial \mathbf{m}}{\partial \rho}\right)^2+\mu_0 M_s^2\left(-\frac{1}{2}k_{\mathrm{E}}m_z^2+\frac{1}{2}k_{\mathrm{H}}m_y^2\right)  \\
& \qquad -\mu_0 M_s\mathbf{m}\cdot\mathbf{H}_z+\mathcal{E}_{\mathrm{iDMI}},
\end{split}
\end{equation}
in which $A$ is the exchange stiffness, $\mu_0$ is the permeability of vacuum,
$M_s$ is the saturation magnetization
and $\mathbf{H}_z=H_z \mathbf{e}_z$ is the external magnetic field along the easy axis 
with the strength $H_z$. 
At last, $k_{\mathrm{E}}(k_{\mathrm{H}})$ denotes the total anisotropy coefficient
along the easy (hard) axis of the central FM layer, 
namely $k_{\mathrm{E}}=k_1 + (D_x-D_z)$ and $k_{\mathrm{H}}=k_2 + (D_y-D_x)$
with $k_{1(2)}$ being the crystalline anisotropy coefficient in easy (hard) axis.

The in-plane charge current flows along ``$\mathbf{e}_{\rho}$" with density $j_a$.
As passing through the MH, the charge current splits into two parts.
Suppose $j_{\mathrm{F}}$ ($j_{\mathrm{H}}$) to be the component in FM (HM) layer.
A simple circuit model delivers that $j_{\mathrm{F}}=j_a(t_{\mathrm{F}}+t_{\mathrm{H}})\sigma_{\mathrm{F}}/(t_{\mathrm{F}}\sigma_{\mathrm{F}}+t_{\mathrm{H}}\sigma_{\mathrm{H}})$ and $j_{\mathrm{H}}=j_a(t_{\mathrm{F}}+t_{\mathrm{H}})\sigma_{\mathrm{H}}/(t_{\mathrm{F}}\sigma_{\mathrm{F}}+t_{\mathrm{H}}\sigma_{\mathrm{H}})$, where $t_{\mathrm{F}}$ ($t_{\mathrm{H}}$) and $\sigma_{\mathrm{F}}$ ($\sigma_{\mathrm{H}}$) 
are the thickness and conductivity of the FM (HM) layer, respectively.
For the most common FM metal (Co, Ni, Fe) and HM (Pt, Ta, Ir) materials, 
the conductivity varies from 10 to 20 $\mathrm{(\mu\Omega m)}^{-1}$.
For simplicity, we set $\sigma_{\mathrm{F}}\approx\sigma_{\mathrm{H}}$ thus $j_{\mathrm{F}}=j_{\mathrm{H}}=j_a$.
The charge current component ($j_{\mathrm{H}}$) in HM substrate will induce a spin current
into the FM layer which is polarized in the direction of
``$\mathbf{m}_{\mathrm{p}}\equiv\mathbf{n}\times\mathbf{e}_{\rho}$",
hence generate the SOT. 
On the other hand, in the global Cartesian coordinate system, 
the unit vector of magnetization in the FM layer can be
fully described by its polar angle $\theta$ and azimuthal angle $\phi$, as shown in Fig. \ref{fig1}. 
The resulting local spherical coordinate system is denoted as 
($\mathbf{e}_{\mathbf{m}},\mathbf{e}_{\theta},\mathbf{e}_{\phi}$).
Then $\mathbf{m}_{\mathrm{p}}$ can be decomposed as
\begin{equation}\label{mp_expression_in_sperical}
	\mathbf{m}_{\mathrm{p}}=p_{\mathbf{m}}\mathbf{e}_{\mathbf{m}}+p_{\theta}\mathbf{e}_{\theta}+p_{\phi}\mathbf{e}_{\phi}.
\end{equation}
Base on all these preparations, the Lagrangian density $\mathcal{L}$ and dissipation functional density
$\mathcal{F}$ of this magnetic system can be expressed as
\begin{equation}\label{Lagrangian_density}
\frac{\mathcal{L}}{\mu_0 M_s^2}=-\frac{\cos\theta}{\gamma M_s}\frac{\partial \phi}{\partial t}-\frac{B_J\phi}{\gamma M_s}\frac{\partial(\cos\theta)}{\partial \rho}+\frac{H_{\mathrm{FL}}}{M_s}p_{\mathbf{m}}-\frac{\mathcal{E}_0}{\mu_0 M_s^2},
\end{equation}
and
\begin{equation}\label{dissipation_functional_density}
\frac{\mathcal{F}}{\mu_0 M_s^2}=\frac{\alpha}{2\gamma M_s}\left\{\left[\frac{\partial}{\partial t}-\frac{\beta B_J}{\alpha}\frac{\partial}{\partial\rho}\right]\mathbf{m}\right\}^2-\frac{H_{\mathrm{ADL}}}{M_s}\left(\mathbf{m}\times\mathbf{m}_{\mathrm{p}}\right)\cdot\frac{\partial\mathbf{m}}{\partial t},
\end{equation}
in which $\gamma=\mu_0\gamma_e$ with $\gamma_e$ being the electron gyromagnetic ratio,
$B_J= \mu_\mathrm{B} P j_a/(e M_s)$ with $e,\mu_\mathrm{B}$ being respectively 
the absolute value of electron charge
and Bohr magneton, $P$ is the spin polarization of $j_{\mathrm{F}}$,
$\alpha$ is the Gilbert damping constant, $\beta$ is the dimensionless coefficient
describing the relative strength of the nonadiabatic STT over the adiabatic one,
at last $H_{\mathrm{FL}}$ and $H_{\mathrm{ADL}}$ denotes the strength of field-like (FL)
and anti-damping-like (ADL) SOT, respectively.

The dynamics of magnetization in the central FM layers of MHs is then described
by the Lagrangian-Rayleigh equation
\begin{equation}\label{Eular_Lagrangian_equation}
	\frac{\mathrm{d}}{\mathrm{d}t}\left(\frac{\delta\mathcal{L}}{\delta\dot{X}}\right)-\frac{\delta\mathcal{L}}{\delta X}+\frac{\delta\mathcal{F}}{\delta X}=0,
\end{equation}
in which $X$ is any related local or collective coordinate.
In particular, when $X=\theta$ and $\phi$ (the most common local coordinates), 
the resulting two equations can be combined to recover the familiar Landau-Lifshitz-Gilbert equation
\begin{equation}\label{LLG_vector}
	\frac{\partial \mathbf{m}}{\partial t}=-\gamma\mathbf{m}\times\mathbf{H}_{\mathrm{eff}}+\alpha\mathbf{m}\times\frac{\partial \mathbf{m}}{\partial t}+\mathbf{T}_{\mathrm{STT}}+\mathbf{T}_{\mathrm{SOT}}, 
\end{equation}
where $\mathbf{H}_{\mathrm{eff}}=-(\mu_0 M_s)^{-1}\delta\mathcal{E}_0/\delta\mathbf{m}$,
and
\begin{equation}\label{STT_definition}
\mathbf{T}_{\mathrm{STT}}= B_J\frac{\partial \mathbf{m}}{\partial \rho}-\beta B_J \mathbf{m}\times\frac{\partial \mathbf{m}}{\partial \rho}, 
\end{equation}
as well as
\begin{equation}\label{SOT_definition}
\mathbf{T}_{\mathrm{SOT}}=-\gamma H_{\mathrm{FL}}\mathbf{m}\times\mathbf{m}_{\mathrm{p}} -\gamma H_{\mathrm{ADL}}\mathbf{m}\times\left(\mathbf{m}\times\mathbf{m}_{\mathrm{p}}\right).
\end{equation}
However, $\theta(\rho,t)$ and $\phi(\rho,t)$ vary from point to point, 
hence generating a huge number of degrees of freedom. 
To obtain collective behaviors of magnetization system, LB-CCMs are adopted which need preset ansatz.
In the beginning of next section, we will define aqequate ansatz for topologically nontrivial 360DWs and 
introduce several typical trial profiles which contain reasonable collective coordinates.
Based on them, a set of dynamical equations can be obtained,
which lays the foundation of our work in this paper.

\section{III. Results}\label{Section_Results}
\subsection{III.A Adequate ansatz for topologically nontrivial 360DWs}\label{Section_Results_Typical_360DW_ansatz}
As we mentioned above, earlier studies confirm that in Q1D MHs if the long-range component of 
magnetostatics is neglected, then an external field along the easy axis is crucial for forming a 360DW.
Accordingly, an analytical profile of static 360DWs has been provided based on the requirement
that at equilibrium the $\mathbf{e}_{\theta}$ component of $\mathbf{H}_{\mathrm{eff}}$ disappears\cite{Muratov_JAP_2008}.
In this solution the azimuthal angle takes a fixed value while the polar angle changes monotonously from 
0 to $\pi$ as $\rho$ runs from one end of MH to the wall center and then decreases back to 0
as $\rho$ goes further to the other end.
This nonmonotonic behavior comes from the consideration that polar angles are defined in
spherical coordinate system thus can not exceed $\pi$.
However, we would like to point out that: 
\emph{a 360DW defined like this must be a topologically trivial one}.
In PMA (IPMA) systems, this corresponds to a ``$\uparrow\rightarrow\downarrow\rightarrow\uparrow$"
(``$\rightarrow$$\uparrow$$\leftarrow$$\uparrow$$\rightarrow$") type
wall which first rotates half a circle and then returns back, thus leading to $\mathcal{W}_{\mathrm{1D}}=0$.

One possible remedy is to add a fixed value $\pi$ to the azimuthal angle when the polar angle 
crosses the South Pole. 
In principle this new set of polar and azimuthal is indeed the real spherical coordinates that realizes a topologically nontrivial ``$\uparrow\rightarrow\downarrow\leftarrow\uparrow$"
(``$\rightarrow$$\uparrow$$\leftarrow$$\downarrow$$\rightarrow$") type wall
for PMA (IPMA) systems, however it will artificially bring 
a discontinuity point in exchange energy.
For the convenience of comparisons below, we denote them as $\vartheta_{\mathrm{real}}$ and $\varphi_{\mathrm{real}}$.
To remove the artificial discontinuity, we propose a monotonically increasing ``0 to $2\pi$" polar 
angle profile meanwhile keep the azimuthal angle a constant value which are defined as
$\vartheta_{\mathrm{ansatz}}$ and $\varphi_{\mathrm{ansatz}}$. 
We focus on the ``$\pi$ to $2\pi$" part since this is the main region where differences occur.
Obviously, we have
\begin{equation}\label{theta_phi_real_ansatz_connection}
\vartheta_{\mathrm{real}}=2\pi-\vartheta_{\mathrm{ansatz}},\quad \varphi_{\mathrm{real}}=\varphi_{\mathrm{ansatz}}+\pi,
\end{equation}
and they lead to the same magnetization component as follows
\begin{equation}\label{theta_phi_real_ansatz_mxyz}
\begin{split}
& \qquad \mathrm{real\ spherical}\qquad\quad \mathrm{ansatz} \\
m_x:&\quad\sin\vartheta_{\mathrm{real}}\cos\varphi_{\mathrm{real}}\equiv\sin\vartheta_{\mathrm{ansatz}}\cos\varphi_{\mathrm{ansatz}}  \\
m_y:&\quad\sin\vartheta_{\mathrm{real}}\sin\varphi_{\mathrm{real}}\equiv\sin\vartheta_{\mathrm{ansatz}}\sin\varphi_{\mathrm{ansatz}}  \\
m_z:&\quad\cos\vartheta_{\mathrm{real}}\quad\quad\quad \equiv\cos\vartheta_{\mathrm{ansatz}}.
\end{split}
\end{equation}
In addition, the polar and azimuthal profiles in our proposal are not bothered by discontinuities
in continuous Heisenberg exchange interaction, meanwhile provide $\mathcal{W}_{\mathrm{1D}}=+1$.
Based on these facts, we conclude that
a 360DW profile with a monotonically increasing ``0 to $2\pi$" polar angle and a constant azimuthal angle
should be an adequate ansatz for 360DWs.

In this work, we use three trial profiles of 360DWs to explore their chirality preference and
current-driven dynamics. The first one is inspired by
the work of Muratov in 2008\cite{Muratov_JAP_2008}, but has been generalized to $\left[0,2\pi\right)$
as we proposed above. By introducing the ``traveling coordinate"
$\xi\equiv\frac{\rho-q(t)}{\Delta(t)}$ where $q(t)$ and $\Delta(t)$ are respectively the center position 
and width of the 360DW, it can be written as
\begin{equation}\label{1st_ansatz}
\vartheta=2\cot^{-1}\left[\sqrt{\frac{h}{1+h}}\sinh\left(-\sqrt{1+h}\xi\right)\right],\quad \phi(\mathbf{r},t)=\varphi(t),
\end{equation}
in which $h\equiv\frac{H_z}{k_{\mathrm{E}}M_s}$ and the ``$\cot^{-1}$" function takes the range of 0 to $\pi$. 
Note that Eq. (\ref{1st_ansatz}) is accurate in the absence of driving current. When in-plane currents
are applied, this solution becomes an approximation since it may not hold everywhere but it does
grasp the main features of dynamical 360DWs. In particular, Eq. (\ref{1st_ansatz}) clearly ascertains the 
conclusion that in the absence of external magnetic fields along the easy axis (i.e. $h=0$), 
$\vartheta$ keeps a constant value thus 360DWs disappear.
Also, we have two other options.
The second trial profile is directly generalized from the Walker ansatz, which reads
\begin{equation}\label{2nd_ansatz}
\vartheta=4\tan^{-1}e^{\xi},\quad \phi(\mathbf{r},t)=\varphi(t),
\end{equation}
and the third one is 
\begin{equation}\label{3rd_ansatz}
\vartheta=\dfrac{2\pi}{1+e^{-\xi}},\quad \phi(\mathbf{r},t)=\varphi(t).
\end{equation}
Obviously, the latter two do not depend on $h$, thus can not be rigorous even in the absence of driving currents.
However, due to their mathematical simplicity, they can be used as references.
In particular when $h=1$ ($h=\pi^2/16$), $\mathrm{d}\vartheta/\mathrm{d}\xi$ at $\xi=0$
in Eq. (\ref{1st_ansatz}) coincides with that of Eq. (\ref{2nd_ansatz}) [Eq. (\ref{3rd_ansatz})].
The corresponding $\vartheta$ profiles are plotted in Fig. 2. 
Also, curves with $h=0.1$ and $h=5$ have been appended 
to illustrate the dependence of polar angle profile in Eq. (\ref{1st_ansatz}) on $h$:
as $h$ increases the effective width [not the parameter $\Delta(t)$] of 360DW is compressed.

\begin{figure} [h] 
	\centering
	\includegraphics[width=0.42\textwidth]{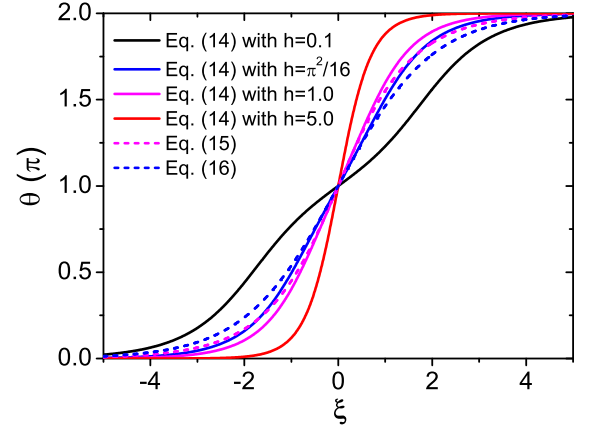}
	\caption{(Color online) Trial polar angle profiles in 
		Eq. (\ref{1st_ansatz}) - Eq. (\ref{3rd_ansatz}). Four solid curves are those from 
	    Eq. (\ref{1st_ansatz}) with different $h$, while the magenta (blue) dashed curve
        shows Eq. (\ref{2nd_ansatz}) [Eq. (\ref{3rd_ansatz})].}\label{fig2}
\end{figure}

\subsection{III.B Dynamical equations}\label{Section_Results_Dynamical_equation_sets}

\begin{table*} [htbp]
	\caption{Summary of parameters in Eq. (\ref{Dynamical_equation_set_general}):
		definitions and values based on the three trial profiles in Eqs. (\ref{1st_ansatz}) to (\ref{3rd_ansatz}).
		First to fifth rows: Integrals $I_1$ to $I_5$. 
		Last row: Parameter $\lambda$.}
	\renewcommand\arraystretch{1.8}
	\begin{tabular} {c|c|c|c|c}
		\hline
		\hline
		Parameter:  & Definition & Value on Eq. (\ref{1st_ansatz})  & Value on Eq. (\ref{2nd_ansatz}) & Value on Eq. (\ref{3rd_ansatz})   \\
		\hline
		$I_1$: & $\Delta\int_{0}^{2\pi}\frac{\partial\vartheta}{\partial\rho}\mathrm{d}\vartheta$ & $4\sqrt{1+h}+2h\ln\frac{\sqrt{1+h}+1}{\sqrt{1+h}-1}$ & $8$ & $\frac{2}{3}\pi^2$   \\
		\hline
		$I_2$:  & $\int_{0}^{2\pi}\sin\vartheta\left(-\xi\right)\mathrm{d}\vartheta$ & $2\ln\frac{\sqrt{1+h}+1}{\sqrt{1+h}-1}$ & $4$ & $2\int_{0}^{2\pi}\frac{1-\cos t}{t}\mathrm{d}t\approx4.8753$  \\
		\hline
		$I_3$: &  $\frac{1}{\Delta}\int_{0}^{2\pi}\frac{\sin^2\vartheta}{\partial\vartheta/\partial\rho}\mathrm{d}\vartheta$ & $4\sqrt{1+h}-2h\ln\frac{\sqrt{1+h}+1}{\sqrt{1+h}-1}$ & $\frac{8}{3}$ &  $\int_{0}^{4\pi}\frac{1-\cos t}{t}\mathrm{d}t\approx3.1144$ \\
		\hline
		$I_4$: & $\Delta\int_{0}^{2\pi}\frac{\partial\vartheta}{\partial\rho}\xi^2\mathrm{d}\vartheta$  & $\frac{8h}{(1+h)^{3/2}}\int_{0}^{+\infty}\frac{\sqrt{1+x^2}\left(\sinh^{-1}x\right)^2}{\left(1+\frac{h}{1+h}x^2\right)^2}\mathrm{d}x$  & $\frac{2}{3}\pi^2$ & $\left(2\pi\right)^2\int_{0}^{+\infty}\frac{x\left(\ln x\right)^2}{\left(1+x\right)^4}\mathrm{d}x\approx8.4870$  \\
		\hline
		$I_5$: & $\int_{0}^{2\pi}\sin\vartheta\cos\vartheta\left(-\xi\right)\mathrm{d}\vartheta$ & \multicolumn{3}{c}{$\frac{I_3}{2}$}   \\
		\hline
		$\lambda$: &   & $I_1-h I_2-I_5$ & $I_1-I_2$ & $\frac{I_1}{2}$ \\
		\hline
		\hline
	\end{tabular}
\end{table*}

\begin{figure} [h] 
	\centering
	\includegraphics[width=0.33\textwidth]{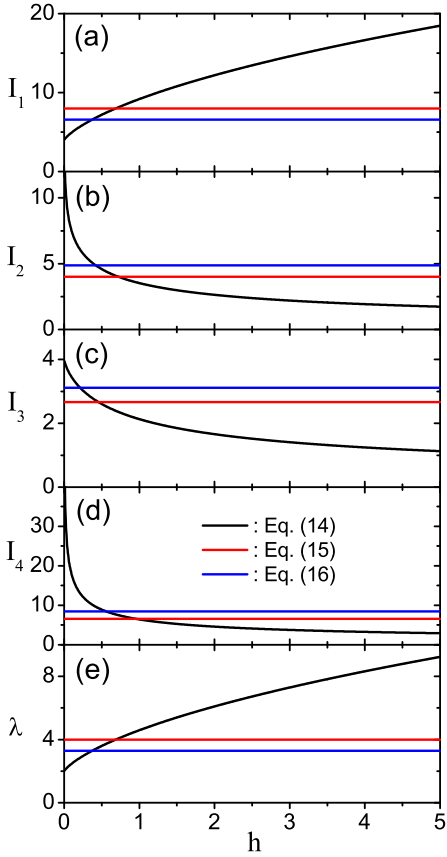}
	\caption{(Color online) Evolution of $I_{1,2,3,4}$ and $\lambda$ as $h$ increases for 
		Eq. (\ref{1st_ansatz}) - Eq. (\ref{3rd_ansatz}). Note that $I_5$ is omitted since it is
		always half of $I_3$. }\label{fig3}
\end{figure}

In all three trial profiles, the wall center position $q(t)$, tilting angle $\varphi(t)$ 
and wall width $\Delta(t)$ are the three collective coordinates.
In Eq. (\ref{Eular_Lagrangian_equation}), by letting $X$ take $q(t)$, $\varphi(t)$, $\Delta(t)$ successively, 
and integrating over the long axis of MHs (i.e. $\int_{-\infty}^{+\infty}\mathrm{d}\rho$), 
a set of dynamic equations can be obtained and expressed in a unified form for both PMA
and IPMA systems:
\begin{subequations}\label{Dynamical_equation_set_general}
\begin{align}
0&=\left(\alpha\dot{q}+\beta B_J\right)-\frac{2\pi}{I_1}\gamma H_{\mathrm{ADL}}\Delta\cdot f(\varphi), \\
\begin{split}
0&=\frac{\alpha}{\gamma M_s} \dot{\varphi}+k_{\mathrm{H}}\sin\varphi\cos\varphi \\
& \quad + \frac{I_2}{I_3}\frac{1}{\gamma M_s}\frac{\dot{\Delta}}{\Delta}+\frac{2\pi}{I_3}\frac{D_{\mathrm{i}}}{\mu_0 M_s^2\Delta}\frac{\mathrm{d}f(\varphi)}{\mathrm{d}\varphi}, 
\end{split} \\
\frac{\alpha I_4}{\gamma M_s}\frac{\dot{\Delta}}{\Delta}&=\frac{I_2}{\gamma M_s}\dot{\varphi}-\left(k_{\mathrm{E}}+k_{\mathrm{H}}\sin^2\varphi\right)I_5-k_{\mathrm{E}}I_2 h +\lambda\frac{l_0^2}{\Delta^2},
\end{align}
\end{subequations}
where an overdot means $\partial/\partial t$ and $l_0=\sqrt{2A/(\mu_0 M_s^2)}$. 
For PMA systems $f(\varphi)=\cos\varphi$ while for IPMA systems $f(\varphi)=-\sin\varphi$. 
The five integrals ($I_1$ to $I_5$) can be defined in a general way without depending on 
the specific form of trial profiles (see the first column of Table I).
Their values and the parameter $\lambda$ under each profile have been
listed in the last three columns of Table I. 
We also plot them in Fig. 3 as functions of $h$ to show
their evolution as $h$ increases.

Eq. (\ref{Dynamical_equation_set_general}) is the starting point for our investigations on chirality and
current-driven dynamics of 360DWs in Q1D MHs.
Before explicitly solving it, we would like to discuss its qualitative properties first.
In the dynamical equations for 180DWs, the iDMI, FL-SOT and ADL-SOT are all present.
However in Eq. (\ref{Dynamical_equation_set_general}), the FL-SOT disappears. 
This can be understood based on the mathematical form of SOTs in Eq. (\ref{SOT_definition}).
The main difference lies in the fact that the FL-term is linear to the magnetization 
$\mathbf{m}$ while the ADL-term is quadratic (thus is nonlinear).
When integrating over the whole strip, the constant ``$-\gamma H_{\mathrm{FL}}\mathbf{m}_{\mathrm{p}}$"
factor can be brought up, leaving ``$\mathbf{m}$" to be integrated over a full circle
thus canceled out.
However, this procedure fails for the ADL-term since 
the constant ``$-\gamma H_{\mathrm{ADL}}\mathbf{m}_{\mathrm{p}}$" factor can not be brought up there.
This explains the presence (absence) of $H_{\mathrm{ADL}}$ ($H_{\mathrm{FL}}$) in Eq. (\ref{Dynamical_equation_set_general}).
Similar analysis can be made to explain the presence of both $H_{\mathrm{ADL}}$ and
$H_{\mathrm{FL}}$ in 180DW case.
Furthermore, a general rule can be summarized as follows: 
\emph{When dealing with current-driven dynamical equations of magnetic domain walls, 
only for ``$2n\pi$" walls $H_{\mathrm{FL}}$ disappears; 
otherwise $H_{\mathrm{ADL}}$ and $H_{\mathrm{FL}}$ coexist.}
Parallel discussions can be performed to the anisotropic field proportional to $k_{\mathrm{H}}$,
which profoundly affect the dynamic behaviors of 360DWs.
We will revisit this issue in Section 3.4.1.

\subsection{III.C iDMI-induced chirality for static 360DWs}\label{Section_Results_chirality}
By first choosing the easy-axis-oriented single-domain state as reference,
and then integrating over the Q1D MH, the ``renormalized magnetic energy" $E_0^{\mathrm{re}}$
of the central FM layer reads
\begin{equation}\label{relative_magnetic_energy}
\begin{split}
\frac{E_0^{\mathrm{re}}}{\mu_0 M_s^2 S}&=\frac{I_1}{2}\frac{l_0^2}{\Delta}+\left[\frac{I_3}{2}\left(k_{\mathrm{E}}+k_{\mathrm{H}}\sin^2\varphi\right)+k_{\mathrm{E}}I_2 h\right]\Delta  \\
& \quad +\frac{2\pi D_{\mathrm{i}}f(\varphi)}{\mu_0 M_s^2},
\end{split}
\end{equation}
where $S$ is the cross section of the central FM layer.
Combing with Eq. (\ref{Dynamical_equation_set_general}) for $j_a=0$ (thus $H_{\mathrm{ADL}}=0$ and $B_J=0$), 
the chirality preference of static 360DWs can be analyzed.

\subsubsection{iDMI is absent}
First we review the simplest case where the iDMI is absent ($D_{\mathrm{i}}=0$).
Physically this corresponds to MHs with normal substrates. 
Then the dynamical equations, as well as the renormalized magnetic energy for PMA and IPMA systems 
are the same.
Since $H_{\mathrm{ADL}}=0$ and $B_J=0$, Eq. (\ref{Dynamical_equation_set_general}a)
provides $\dot{q}=0$ meaning that the 360DW keeps static.
A static wall also requires that $\dot{\varphi}=0$ and $\dot{\Delta}=0$.
Putting them into Eq. (\ref{Dynamical_equation_set_general}b), one has $\sin 2\varphi=0$ which
means $\varphi=\frac{n\pi}{2}$. However, Eq. (\ref{relative_magnetic_energy}) clearly
tells us that only $\varphi=n\pi$ (i.e. $\sin\varphi=0$) minimizes $E_0^{\mathrm{re}}$.
Therefore, in the absence of iDMI, 360DWs should be N\'{e}el type, but have no chirality
preference. At last, Eq. (\ref{Dynamical_equation_set_general}c) provides the static wall width
$\Delta_0$ as
\begin{equation}\label{static_wall_width_no_iDMI}
\Delta_0=\frac{l_0}{\sqrt{k_{\mathrm{E}}}}\sqrt{\frac{\lambda}{I_2 h+I_5}}.
\end{equation}
Note that $\Delta_0$ should not be obtained from the direct minimization of the first two terms
in Eq. (\ref{relative_magnetic_energy}) since the result may not satisfy the dynamical equations.
This argument also holds when iDMI appears.

\subsubsection{PMA systems with iDMI}
For PMA systems, $f(\varphi)=\cos\varphi$. 
The combination of Eq. (\ref{Dynamical_equation_set_general}b) and the static requirement
($\dot{\varphi}=0$ and $\dot{\Delta}=0$) leads to
\begin{subequations}\label{phi_solutions_PMA_with_iDMI}
    \begin{empheq}[left=\empheqlbrace]{align}
    \mathrm{case\; (a):}\quad\,\sin\varphi&=0 \qquad \mathrm{or}\\
    \mathrm{case\; (b):}\quad\cos\varphi&=\frac{2\pi D_{\mathrm{i}}}{k_{\mathrm{H}}I_3 \mu_0 M_s^2\Delta}
    \end{empheq}
\end{subequations}
To determine which solution provides the real tilting angle, we must compare the corresponding
``renormalized magnetic energy" in Eq. (\ref{relative_magnetic_energy}).
For case (a), $\sin\varphi=0\Leftrightarrow \varphi=n\pi$. 
However, the existence of iDMI [the last term in Eq. (\ref{relative_magnetic_energy})]
breaks the two-fold degeneracy of $E_0^{\mathrm{re}}$ upon azimuthal angle.
To minimize $E_0^{\mathrm{re}}$, one must have 
\begin{equation}\label{phi_solution_a_PMA_with_iDMI}
\cos\varphi=-\mathrm{sgn}\left(D_{\mathrm{i}}\right),
\end{equation}
where ``sgn" denotes the sign function. Correspondingly in this case the renormalized magnetic energy becomes
\begin{equation}\label{relative_magnetic_energy_phi_solution_a_PMA_with_iDMI}
\frac{\left(E_0^{\mathrm{re}}\right)_a}{\mu_0 M_s^2 S}=\frac{I_1}{2}\frac{l_0^2}{\Delta}+k_{\mathrm{E}}\left(\frac{I_3}{2}+I_2 h\right)\Delta-\frac{2\pi |D_{\mathrm{i}}|}{\mu_0 M_s^2}.
\end{equation}
For case (b), direct calculation yields
\begin{equation}\label{relative_magnetic_energy_phi_solution_b_PMA_with_iDMI}
\begin{split}
\frac{\left(E_0^{\mathrm{re}}\right)_b}{\mu_0 M_s^2 S}&=\frac{I_1}{2}\frac{l_0^2}{\Delta}+k_{\mathrm{E}}\left(\frac{I_3}{2}+I_2 h\right)\Delta \\
& \quad +\left[\frac{\left(2\pi D_{\mathrm{i}}\right)^2}{2k_{\mathrm{H}}I_3\left(\mu_0 M_s^2\right)^2\Delta}+\frac{I_3}{2}k_{\mathrm{H}}\Delta\right].
\end{split}
\end{equation}
Obviously for any positive $\Delta$, we always have $\left(E_0^{\mathrm{re}}\right)_a<\left(E_0^{\mathrm{re}}\right)_b$. 
Therefore for PMA systems, Eq. (\ref{phi_solution_a_PMA_with_iDMI}) provides
the real azimuthal angle of 360DWs, which presents definite chirality uniquely determined by iDMI. 
This can be understood more intuitively from the perspective of effective fields.
For PMA systems, the iDMI energy density in Eq. (\ref{E_iDMI_expression_general})
leads to the following effective field
\begin{equation}\label{iDMI_effective_field_PMA}
\mathbf{H}_{\mathrm{i}}=-\frac{1}{\mu_0}\frac{\delta\mathcal{E}_{\mathrm{iDMI}}}{\delta\mathbf{M}}=-\frac{2D_{\mathrm{i}}}{\mu_0 M_s}\left[\left(\frac{\partial m_x}{\partial x}\right)\mathbf{e}_z-\left(\frac{\partial m_z}{\partial x}\right)\mathbf{e}_x\right].
\end{equation}
Clearly the $x-$component leads to the chirality of 360DWs.

At last, by putting Eq. (\ref{phi_solution_a_PMA_with_iDMI}) into Eq. (\ref{Dynamical_equation_set_general}c),
the static wall width is found to be the same as that in Eq. (\ref{static_wall_width_no_iDMI}).
For the first trial profile [see Eq. (\ref{1st_ansatz})], 
$I_2$, $I_5$ and $\lambda$ are all functions of $h$. 
One can easily check that $\sqrt{\lambda/(I_2 h+I_5)}=1$, which means that 
$\Delta_0$ is independent on $h$. This is reasonable since in this profile
$\Delta$ and $h$ appear together and are independent variables.
While for the other two profiles,  $I_2$, $I_5$ and $\lambda$ are constants.
Then the wall width will be compressed when an external field in the easy axis appears, which
is also reasonable since $h$ is absent in these two.

\subsubsection{IPMA systems with iDMI}
For IPMA systems, parallel discussions can be performed.
For brevity, we only list the main results here.
Since $f(\varphi)=-\sin\varphi$, the static condition then provides
\begin{subequations}\label{phi_solutions_IPMA_with_iDMI}
	\begin{empheq}[left=\empheqlbrace]{align}
	\mathrm{case\; (a'):}\quad\cos\varphi&=0 \qquad \mathrm{or}\\
	\mathrm{case\; (b'):}\quad\,\sin\varphi&=\frac{2\pi D_{\mathrm{i}}}{k_{\mathrm{H}}I_3 \mu_0 M_s^2\Delta}
	\end{empheq}
\end{subequations}
For case (a'), the iDMi-induced chirality selects $\sin\varphi=\mathrm{sgn}\left(D_{\mathrm{i}}\right)$.
However after simple calculation, it is easy to find that the renormalized magnetic energy
in case (b') is lower than that in case (a').
Therefore the correct static azimuthal angle for IPMA systems
should be the one in Eq. (\ref{phi_solutions_IPMA_with_iDMI}b).
Also, the wall acquires definite chirality determined by the iDMI.

Again, putting Eq. (\ref{phi_solutions_IPMA_with_iDMI}b) back into Eq. (\ref{Dynamical_equation_set_general}c),
the static wall width for IPMA systems is
\begin{equation}\label{static_wall_width_IPMA_with_iDMI}
\Delta'_0=\frac{l_0}{\sqrt{k_{\mathrm{E}}}}\sqrt{\frac{\lambda}{I_2 h+I_5}}\sqrt{1-\frac{k_{\mathrm{H}}I_5}{\lambda}\left(\frac{2\pi D_{\mathrm{i}}}{k_{\mathrm{H}}I_3 \mu_0 M_s^2 l_0}\right)^2}.
\end{equation}
Compared the above result with that in PMA case [see Eq. (\ref{static_wall_width_no_iDMI})],
a quadratic correction term of $D_{\mathrm{i}}$ appears.
For the second and third trial profiles, it does not change the dependence trend of wall width on $h$.
However, for the first profile a problem emerges since now $\Delta$ depends on $h$.
This means that in IPMA systems, Eq. (\ref{1st_ansatz}) is not as good as it is in PMA systems.
The reason lies in the fact that in IPMA systems the hard axis is along $y-$axis (rather than $x-$axis)
since it is the thinnest direction of the strip thus has the largest demagnetization factor.
Despite this, Eq. (\ref{1st_ansatz}) does grasp the main features of 360DWs in IPMA systems
and should be a good ansatz to explore their statics and dynamics.

\subsection{III.D Current-driven 360DW dynamics}\label{Section_Results_dynamics}
When in-plane currents are applied, the 360DWs will be driven to propagate along
the long axis of MHs. 
Generally, in Eq. (\ref{Dynamical_equation_set_general}a) the presence of ``$H_{\mathrm{ADL}}$" term
will change the wall's mobility from the pure STT-driven result by means of $f(\varphi)$. 
To acquire the time-evolution of $\varphi$, Eq. (\ref{Dynamical_equation_set_general}b) and
(\ref{Dynamical_equation_set_general}b) provides
\begin{equation}\label{phi_dynamics_general}
\frac{\mathrm{d}\left(2\varphi+\kappa\right)}{\Gamma-\dfrac{4\pi D_{\mathrm{i}}}{k_{\mathrm{H}}I_3\mu_0 M_s^2\Delta}\dfrac{\mathrm{d}f(\varphi)}{\mathrm{d}\varphi}-\sin\left(2\varphi+\kappa\right)}=\chi \mathrm{d}t,
\end{equation}
with
\begin{equation}\label{kappa_Gamma_chi_definitions}
	\begin{split}
	\kappa&=\arctan\frac{I_2 I_5}{\alpha I_3 I_4}, \\
	\Gamma&=\frac{2 I_2}{\alpha k_{\mathrm{H}} I_3 I_4}\left[\frac{I_5 k_{\mathrm{H}}}{2}+k_{\mathrm{E}}\left(I_2 h+I_5\right)-\lambda \frac{l_0^2}{\Delta^2}\right], \\
	\chi&=k_{\mathrm{H}}\gamma M_s\left[\alpha+\frac{\left(I_2\right)^2}{\alpha I_3 I_4}\right]^{-1}>0.
	\end{split}
\end{equation}
This is the fundamental equation when dealing with current-driven 360DW dynamics.

\subsubsection{iDMI is absent}
First we consider the simplest case where the iDMI is absent ($D_{\mathrm{i}}=0$), which corresponds to 
a 360DW residing in a MH with a normal substrate. Note that at this moment the SOT is also absent.
Now Eq. (\ref{phi_dynamics_general}) can be directly integrated out and the
result depends on the value of $\Gamma$.

When $|\Gamma|<1$,  
\begin{equation}\label{phi_solution_at_Gamma_lt_1_without_iDMI}
\tan\left(\varphi+\frac{\kappa}{2}\right)=\frac{1}{\Gamma}-\frac{\sqrt{1-\Gamma^2}}{\Gamma}\frac{C_1 e^{\sqrt{1-\Gamma^2}\chi t}+1}{C_1 e^{\sqrt{1-\Gamma^2}\chi t}-1},
\end{equation}
with 
\begin{equation*}
C_1=\frac{\Gamma\tan\left(\varphi_0+\dfrac{\kappa}{2}\right)-1-\sqrt{1-\Gamma^2}}{\Gamma\tan\left(\varphi_0+\dfrac{\kappa}{2}\right)-1+\sqrt{1-\Gamma^2}},
\end{equation*}
and $\varphi=\varphi_0$ at $t=0$.
Obviously when $t\rightarrow +\infty$ the azimuthal angle approaches the following value
\begin{equation}\label{phi_final_at_Gamma_lt_1_without_iDMI}
\varphi_{\infty} =\arctan\left(\frac{1}{\Gamma}-\frac{\sqrt{1-\Gamma^2}}{\Gamma}\right)-\frac{\kappa}{2}.
\end{equation}
This means that in this case the 360DW will eventually fall into the ``steady-flow" mode.
By letting $\dot{\varphi}=0$ and $\dot{\Delta}=0$, we know that the wall propagates with a constant velocity
$-\beta B_J/\alpha$ and a fixed width
\begin{equation}\label{wall_width_at_Gamma_lt_1_without_iDMI}
\Delta\left(\varphi_{\infty}\right)=\frac{l_0}{\sqrt{k_{\mathrm{E}}}}\sqrt{\frac{\lambda}{I_2 h+I_5+\left(k_{\mathrm{H}}/k_{\mathrm{E}}\right)I_5\sin^2\varphi_{\infty}}}.
\end{equation}

When $|\Gamma|>1$, 
\begin{equation}\label{phi_solution_at_Gamma_gt_1_without_iDMI}
\varphi=\arctan\left[\frac{\sqrt{\Gamma^2-1}}{\Gamma}\tan\left(\frac{\sqrt{\Gamma^2-1}}{2}\chi t+C_2\right)+\frac{1}{\Gamma}\right]-\frac{\kappa}{2},
\end{equation}
with 
\begin{equation*}
C_2=\arctan\frac{\Gamma\tan\left(\varphi_0+\dfrac{\kappa}{2}\right)-1}{\sqrt{\Gamma^2-1}}.
\end{equation*}
Now the azimutial angle rotates periodically with the period 
\begin{equation}\label{phi_period_at_Gamma_gt_1_without_iDMI}
T_0=\frac{4\pi}{\chi\sqrt{\Gamma^2-1}},
\end{equation}
which means that the 360DW takes a ``precessional-flow" mode with the same constant velocity
$-\beta B_J/\alpha$ and a periodically changing width.

It is worth noting that for a certain 360DW as the current density increases the wall always takes a specific
mode (either steady-flow or precessional-flow) rather than going through a process of mode change,
which is quite different from the commonly studies 180DWs. 
This is the direct consequence of the ``full-circle" topology that 360DWs hold.
Similar with the discussions in Sec. 3.2, for 180DWs [or other ``$(2n+1)\pi$" walls] 
the incomplete cancellation over a half-circle rotation of $\mathbf{m}$ 
leads to the appearance of ``$k_{\mathrm{H}}\sin2\varphi$" term, and then results in
the famous ``Walker breakdown" process.
However for 360DWs (or other ``$2n\pi$" walls), 
the full cancellation of anisotropic field leads to the absence of ``$k_{\mathrm{H}}\sin2\varphi$" term
in Eq. (\ref{Dynamical_equation_set_general}), thus results in the ``fixed mode" behavior.
Interestingly, both modes share the same wall mobility which is equal to that in steady-flow mode of 180DWs.
This explains nearly all existing numerical observations before 360DWs change to 
other magnetic solitons (for example vortices) under too high currents \cite{Ross_PRB_2010,HuJingGuo_AIPAdvances_2015,Xibin_srep_2017}.

Furthermore, we provide the sufficient but non-necessary condition for the steady flow of 360DWs.
Under the presupposition $|\Gamma|<1$, by putting the wall width [see Eq. (\ref{wall_width_at_Gamma_lt_1_without_iDMI})] back into the definition 
of $\Gamma$ [see Eq. (\ref{kappa_Gamma_chi_definitions})], we obtain
\begin{equation}\label{Gamma_lt_1_simplified_without_iDMI}
\Gamma=\frac{I_2 I_5}{\alpha I_3 I_4}\cos 2\varphi_{\infty}.
\end{equation}
Thus the sufficient but non-necessary condition for $|\Gamma|<1$ should be $I_2 I_5/(\alpha I_3 I_4)<1$,
which corresponds to $\alpha>\alpha_c\equiv I_2 I_5/(I_3 I_4)$.
For the second and third trial profiles, one has $\alpha_c=3/\pi^2\approx 0.304$ and $\alpha_c=0.287$, respectively.
While for the first profile, $\alpha_c$ is the function of $h$. 
We have plotted their dependence on $h$ in Fig. 4. 
One can clear see that for all three cases $\alpha_c$ has a upper limit $3/\pi^2$
even when $h$ increases to 5 which is a quite high value in real experiments.
In many MHs, existing measurements show that the effective damping in FM strips
is enhanced from 0.001-0.01 to 0.3-0.9\cite{Pizzini_EPL_2016,Marrows_PRB_2018}. 
This guarantees that experimentally 360DWs should take the steady-flow mode.
As for precessional flow, since the explicit form of wall width is hard to obtain, thus it is difficult 
to obtain the definite range of its existence.
However, from the above discussion we can reasonably infer that for sufficient small $\alpha$,
360DWs precess.
This prediction needs to be verified by future experiments and numerical simulations.

\begin{figure} [h] 
	\centering
	\includegraphics[width=0.4\textwidth]{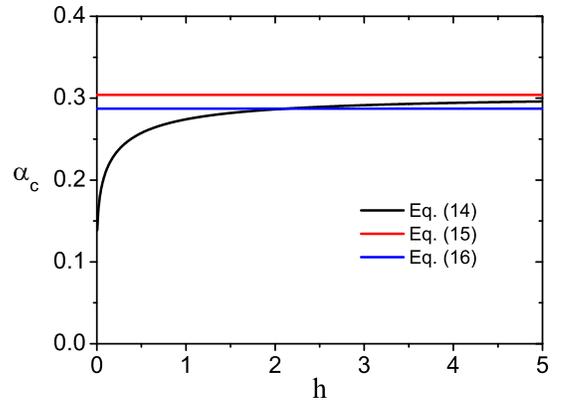}
	\caption{(Color online) Dependence of $\alpha_c$ on $h$ based on 
		Eq. (\ref{1st_ansatz}) - Eq. (\ref{3rd_ansatz}). }\label{fig4}
\end{figure}

\subsubsection{iDMI and ADL-SOT are present}
Next we study the effects of iDMI and ADL-SOT on the current-driven dynamics of 360DWs in
MHs with HM substrates.
In the presence of iDMI, in principle the azimuthal angle $\varphi$ can not be integrated out
explicitly from Eq. (\ref{phi_dynamics_general}).
Recently a phase diagram has been drawn to show how the types of solutions are determined by 
the DMI and the anisotropic parameters\cite{LiZaidong_JMMM_2020}.
However in real MHs, generally the iDMI is weaker than other magnetic interactions,
thus can be reasonably viewed as a small quantity.
Depending on the value of $\Gamma$, different approximate treatments will be used.

When $|\Gamma|<1$ or under the stronger condition $\alpha>\alpha_c$, 
at the lowest level of approximation the 360DW should
eventually propagate like a rigid body with the finial azimtuhal angle $\varphi_{\infty}$, width $\Delta(\varphi_{\infty})$
and velocity
\begin{equation}\label{Gamma_lt_1_360DW_velocity_with_iDMI}
\dot{q}=-\frac{\beta}{\alpha}B_J+\frac{2\pi}{\alpha I_1}\gamma H_{\mathrm{ADL}}\Delta(\varphi_{\infty})\cdot f(\varphi_{\infty}).
\end{equation}
Obviously, the wall mobility is modified by the second term.
However the effect of iDMI is totally submerged since it has been dropped when obtaining
$\varphi_{\infty}$. Note that the form of $\Delta(\varphi_{\infty})$ in 
Eq. (\ref{wall_width_at_Gamma_lt_1_without_iDMI}) is not effected by this dropping.

When $|\Gamma|>1$, the 360DW precesses. In this case for a physical quantity $O$, its time average
\begin{equation}\label{time_average_definition}
\langle O\rangle\equiv \frac{1}{T} \int_{0}^{T}X(t)\mathrm{d}t=\frac{1}{T}\int_{0}^{2\pi}\frac{X}{\dot{\varphi}}\mathrm{d}\varphi
\end{equation}
corresponds to experimental observables, where $T$ is the precession period.
Under the assumption of small iDMI, we calculate the time-averaged wall velocity
$\langle \dot{q}\rangle$.
First the period $T$ is replaced by $T_0$ in Eq. (\ref{phi_period_at_Gamma_gt_1_without_iDMI}).
Then the approximation $(1-x)^{-1}\approx 1+x+x^2$ for $|x|<1$ is used to simplify $1/\dot{\varphi}$
in Eq. (\ref{phi_dynamics_general}) hence the integral in Eq. (\ref{time_average_definition}) can
be calculated. After standard algebra, we have
\begin{equation}\label{Gamma_gt_1_360DW_velocity_with_iDMI}
\langle \dot{q}\rangle=-\frac{\beta}{\alpha}B_J-\eta\frac{\sqrt{\Gamma^2-1}}{\Gamma^3}\gamma H_{\mathrm{ADL}}\frac{D_{\mathrm{i}}}{\mu_0 M_s^2}\frac{4\pi^2\cos\kappa}{\alpha k_{\mathrm{H}}I_1 I_3},
\end{equation}
where $\eta=+1$ ($-1$) for PMA (IPMA) systems.
Clearly, Eq. (\ref{Gamma_gt_1_360DW_velocity_with_iDMI}) provides the effects of both ADL-SOT and iDMI 
to the wall velocity in precessional flows.

\section{IV. Discussions}\label{Section_Discussion}
First, one should note that the premise of all our analytical results is the existence of 360DWs. 
The constant mobility (whether adjusted by iDMI and SOT or not) upon current increase
is the direct manifestation of the wall's ``full-circle" topology.
Accordingly, strong enough external stimuli would destroy the configuration of 360DWs, 
thereby greatly change the mobility of domain walls (not 360DWs any more).
This explains the huge reduction of 360DW mobility under high currents
in existing numerics\cite{Ross_PRB_2010,HuJingGuo_AIPAdvances_2015,Xibin_srep_2017}.

Second, our analytics presented here is based on ``$0$ to $2\pi$" monotonic profiles of polar angle.
If $\vartheta$ is no longer monotonic but its overall change across the wall region keeps $2\pi$
($\mathcal{W}_{\mathrm{1D}}=+1$ still holds), then the results will be unchanged.
In addition, for a 360DW with $\mathcal{W}_{\mathrm{1D}}=-1$ mathematically
its profile can be transfer to that with $\mathcal{W}_{\mathrm{1D}}=+1$, except for
an increase by $\pi$ in the azimuthal angle.
The following procedure is similar to what we have presented in the main text and
will not provide new physics, so we won't repeat it.

At last, topologically the 1D 360DWs in narrow MHs under investigation here are analogous to
the 1D domain wall skyrmions (DWSs) evolved from vertical Bloch lines in wide MHs with PMA\cite{ChengRan_PRB_2019}.
Both magnetic solitons carry integer 1D topological charges ($\mathcal{W}_{\mathrm{1D}}=\pm 1$),
hence should belong to the same topology class.
The effective field of iDMI in that work plays the role of external fields along easy axis here,
therefore is crucial to the formation of 1D DWSs.
The current-driven results here may provide insights for exploring dynamical behaviors of 
1D DWSs under external stimuli.

\section{V. Conclusion}\label{Section_Conclusion}
In this work, the topology, chirality and current-driven dynamics of 360DWs in Q1D MHs 
are systematically investigated.
On one hand, the iDMI uniquely select the chirality of static 360DWs.
On the other hand, the ``full-circle" topology of 360DWs makes them completely different 
from the traditional 180DWs.
For 360DWs, effective fields which are linear to the magnetization have been fully
canceled out and disappear in the dynamical equations.
In particular, the full cancellation of magnetic anisotropic fields directly results in
the absence of ``Walker breakdown"-type process under increasing currents.
In a certain MH, 360DWs will take either steady-flow or precessional-flow mode, depending on
the strength of effective Gilbert damping constant therein.
In MHs with normal substrates, the wall mobility of both modes are the same as that 
in the steady-flow mode of STT-driven propagation of 180DWs.
While in MHs with HM substrates, the mobility will be modified by the ADL-SOT and iDMI.
These results should deepen our understanding of topological solitons in low-dimensional magnetic systems,
meanwhile provide necessary theoretical basis for expanding the application 
of 360DWs in the field of magnetic nanodevices.

\section{Acknowledgement}
M. L. is supported by the National Natural Science Foundation of China (Grants No. 11947023)
and the Project of Hebei Province Higher Educational Science and Technology Program (QN2019309).
J. L. acknowledges supports from Natural Science Foundation for Distinguished Young Scholars of
Hebei Province of China (A2019205310) and from National Natural Science Foundation of China (Grants No. 11374088).

\end{document}